# Electrostatic complexation of spheres and chains under elastic stress

H. Schiessel[1], R. F. Bruinsma[2] and W. M. Gelbart[3]

[1] *Max-Planck-Institute for Polymer Research, Theory Group, POBox 3148, 55021 Mainz, Germany*, [2]*Instituut-Lorentz for Theoretical Physics, Universiteit Leiden, Postbus 9506, 2300 Leiden, The Netherlands*, [3]*Department of Chemistry and Biochemistry, University of California, Los Angeles CA 90095-1569, USA*

ABSTRACT We consider the complexation of highly charged semiflexible polyelectrolytes with oppositely charged macroions. On the basis of scaling arguments we discuss how the resulting complexes depend on the persistence length of the polyelectrolyte, the salt concentration, and the sizes and charges of the chain and the macroions. We study first the case of complexation with a single sphere and calculate the wrapping length of the chain. We then extend our considerations to complexes involving many wrapped spheres and study cooperative effects. The mechanical properties of such a complex under an external deformation are evaluated.

## 1 Introduction

The complexation of polyelectrolytes and oppositely charged macroions is a primary ingredient in biological processes. The non-specific part of the interaction between proteins and DNA is governed by electrostatics. A well-known example of this form of complexation is the association of DNA with oppositely charged octamers of histone proteins, an essential step in chromosomal DNA compaction.[1] Complexation of macroions is also encountered in several technological applications. For instance, the complexation of synthetic polymers with colloidal particles[2,3] and charged micelles[4] is of practical importance for modifying macro-ion solution behavior.

A number of experimental[5,6,7,8] and theoretical studies[9,10,11,12,13] have demonstrated that complexation of highly charged macro-ions is governed by an unusual electrostatics mechanism: *counterion release*. The electrostatic free energy of association of oppositely charged macro-ions is dominated by the entropy increase arising from the release of counterions that had been condensed onto the macro-ions before association. This electrostatic free energy gain must compete with a free energy cost induced by deforming either or both of the macroions so as to bring the fixed macro-ion charges of opposite sign in close contact.

A simple example of this competition, first discussed by Marky and Manning[14], is the association of a charged sphere with an oppositely charged semiflexible chain. If $R$ is the radius of the sphere, $l_P$ the persistence length of the chain (i.e., the bending modulus of the chain equals $k_B T l_P$ with $k_B T$ the thermal energy), and $\mathbf{l}$ the electrostatic free energy gain per unit length of adhesion, then the free energy cost of association is:

$$F(l) \cong \left(\frac{k_B T l_P}{R^2} - \mathbf{l}\right) l + O(l^2) \qquad (1)$$

with $l$ the length of chain wrapped around the sphere. Under conditions where counterion release dominates (i.e., high bare charges), $\mathbf{l}$ is of the order $k_B T/b$ with $b$ the spacing between charges along the chain. According to Eq. (1), complexation starts when $\mathbf{l}$ exceeds $k_B T l_P/R^2$. It would appear reasonable that wrapping continues until the charge of the wrapped part of the chain has compensated the charge of the central sphere. The counterion release mechanism produces a surprise: An analysis [10] for this case found that, for small enough persistence lengths, the chain/sphere complex is *overcharged*: more chain is wrapped on the sphere then required for charge compensation. "Charge-reversal" normally is associated with short-range correlations between the charges,[15] but here it is again due to entropy increase of the counterions.

In the present paper we extend the above analysis to examine complexation of a flexible charged chain placed in a *solution* of oppositely charged macroions, assuming that the complex adopts a "beads-on-a-string" geometry. (This particular geometry is, for example, encountered for DNA/histone complexes at low salt concentrations). A similar system of charged spheres and chains was also recently investigated by Nguyen and Shklovskii.[16] Their study focuses on weakly charged systems where counterions are not important. As we discuss in the conclusion of this paper their system shows nevertheless many features that are characteristic of our system. It should be noted though that such simple models are only useful for a discussion of the generic aspects of complexation of charged linear macromolecules with spherical macroions. For any particular case, the specific aspects of the molecular interactions for that situation must be accounted for.

The central claim of this paper is that whereas complexation of a chain with a single sphere leads to spontaneous overcharging, complexation in a solution of spherical macroions leads to spontaneous *undercharging*, even though both are due to the same counterion release mechanism. The surprising role reversal is reflected in force-extension curves of the kind now routinely measured for long biopolymers. For the case of individual sphere/chain complexes the effect of an external tension *f* can be accounted for by adding, in Eq. (1), a term *fl*. At a critical tension equal to $\mathbf{l} - k_B T l_P/R^2$, the chain-sphere complex dissociates. The measurement of the force-extension curve thus gives information on the adhesion energy per unit length. For the case of a chain under tension in chemical equilibrium with a solution of spherical macro-ions, however, we find that with increasing tension more and more spheres condense on the chain and that the critical tension to add one additional sphere vanishes in the thermodynamic limit of an infinitely long chain. We also consider a chain complexed with a fixed number *N* of spherical macroions under an externally imposed strain. Here one might expect that beads are released with increasing strain in order to have the remaining beads close to their optimal wrapping length – resulting in a "saw-tooth-like" force-extension curve of the type recently encountered for the tension-induced denaturation of the linear macromolecule titin.[17,18] Again, our finding comes as a surprise: The chain simultaneously unwraps *all* the beads in parallel and there is no sequential release.

In the next section we start by reviewing the statistical thermodynamics of complexation of a single sphere with an oppositely charged flexible rod and discuss the mechanism of spontaneous overcharging. Next we treat the complexation and the force-extension behavior of a chain in chemical equilibrium with a solution of free macroions, and then the complexation and force-extension curve of a chain with a fixed number of spheres.

## 2 Complexation of a chain and a single sphere

Consider the case of a single sphere of radius $R$ and charge $eZ$ and a semiflexible rod with a charge per unit length $-e/b$, persistence length $l_P$, length $L \gg R$ and radius $r$. They are both placed in a salt solution characterized by: a Bjerrum length $l_B \equiv e^2/ \varepsilon k_B T$, with $\varepsilon$ the dielectric constant of the solvent; and a Debye screening length $\kappa^{-1} = (8\pi c_s l_B)^{-1/2}$, with $c_s$ the bulk concentration of salt. We will assume salt concentrations such that $\kappa^{-1}$ is large compared to the sphere radius, $\kappa R \ll 1$, but small compared to $L$. The persistence length is assumed large compared with $R$, in contrast to the case of complexation of highly flexible chains with spheres ($R \gg l_B$) studied by Pincus et al.[19] and recently by Nguyen and Shklovskii.[16]

We restrict our study to highly charged chains for which the Manning parameter $\xi \equiv l_B/b$ is much larger than one. In this case $(1-\xi^{-1})L/b \cong L/b$ counterions are condensed on the chain.[20,21] The entropic "confinement" cost is $\Omega k_B T$ per condensed counterion with $\Omega = 2\ln(4\xi\kappa^{-1}/r)$.[22,23] The total entropic electrostatic charging free energy of the chain in this case ($\xi \gg 1$) is then given by

$$F_{chain}(L) \cong \frac{k_B T}{b} \Omega L \qquad (2)$$

On the other hand, the electrostatic charging free energy of a spherical macroion of charge $Z$ is

$$F_{sphere}(Z) \cong \begin{cases} \dfrac{e^2 Z^2}{2\varepsilon R} & \text{for } |Z| < Z_{max} \\ |Z|k_B T \tilde{\Omega}(Z) & \text{for } |Z| \gg Z_{max} \end{cases} \qquad (3)$$

where $\tilde{\Omega}(Z) = 2\ln(|Z|l_B \kappa^{-1}/R^2)$ (cf. Ref. 22) and $Z_{max} \approx R/l_B$. For weakly charged spheres, $|Z| < Z_{max}$, $F_{sphere}$ is the electrostatic charging energy, Eq. (3) $|Z| < Z_{max}$. In the case of highly charged spheres with $|Z| \gg Z_{max}$ most of the counterions are localized close to the sphere with an entropic cost $\tilde{\Omega}(Z) k_B T$ per counterion leading to Eq. 3 for $|Z| \gg Z_{max}$. In that case only a small fraction $Z_{max}/Z$ of counterions is still free. $Z_{max}$ − which is also the effective charge of the sphere − follows a the balance of electrostatic charging energy $l_B Z_{max}^2 / 2R$ and entropy $\tilde{\Omega}(Z) Z_{max}$, and is therefore of the order $\tilde{\Omega} R / l_B$.[24]

We will determine the total free energy of the chain/sphere complex as follows. Assume that a length $l$ of the chain has been wrapped around the sphere. We divide the sphere/chain complex in two parts: the sphere with the wrapped part of the chain − of length $l$ − and the remaining chain of the length $L-l$. The first part − which we will refer to as the "complex" ("*compl*") − has a total charge

$$Z(l) = Z - l/b \tag{4}$$

We estimate the electrostatic free energy $F_{compl}(l)$ of the complex by $F_{sphere}(Z(l))$. Eqs. (3) and (4) then imply that $F_{compl}$ varies quadratically as $(Z - l/b)^2$ for $|Z(l)| < Z_{max}$ whereas it will be approximately linear for $|Z(l)| > Z_{max}$. Note that this procedure neglects higher-order multipole contributions that may play a role for small $Z(l)$. Note also that there is a special length $l_{iso} = bZ$ such that $Z(l_{iso}) = 0$. Simply invoking the principle of charge neutrality would lead one to expect that the total free energy is minimized for $l \cong l_{iso}$.

The total free energy is:

$$F(l) = F_{compl}(l) + F_{chain}(L-l) + F_{compl-chain}(l) + F_{elastic}(l) \tag{5}$$

The first two terms have already been discussed in Eqs. (2) to (4). The third term is the electrostatic free energy of the interaction between the complex and the remainder of the chain. This is of the order:

$$F_{compl-chain}(l) \cong Z^*(l) \ln(kR) \tag{6}$$

where $Z^*(l)$ is the *effective* charge of the complex. For small complex charges with $Z(l) < Z_{max}$ the effective charge obeys $Z^*(l) = Z(l)$; in the opposite case, $Z(l) > Z_{max}$, one has $Z^*(l) = Z_{max}$, thereby making $F_{compl-chain}$ independent of *l*. The final term in Eq. (5) − the elastic free energy − was already discussed in Section 1:

$$F_{elastic}(l) \cong \left( \frac{k_B T \, l_P}{R^2} + f \right) l \tag{7}$$

We will consider separately the two cases $|Z(l)| < Z_{max}$ and $|Z(l)| > Z_{max}$.[25] For the first case we find (for $f = 0$)

$$\frac{F(l)}{k_B T} \cong \frac{l_B}{2R}\left(Z - \frac{l}{b}\right)^2 + Al/b + const \qquad l_{min} < l < l_{max} \qquad (8)$$

where

$$A = \frac{l_p b}{R^2} - \ln(kR) - \Omega \qquad (9)$$

The validity limits $l_{min}$ and $l_{max}$ are found by equating $Z(l)$ with $\pm Z_{max}$: $l_{min} = l_{iso} - bZ_{max}$ and $l_{max} = l_{iso} + bZ_{max}$. For the second case, $|Z(l)| > Z_{max}$,

$$\frac{F(l)}{k_B T} \cong B^{\mp} l/b + const \qquad (10)$$

with

$$B^{\mp} = \frac{l_p b}{R^2} - \Omega \mp \tilde{\Omega} \qquad (11)$$

The "−" sign refers to the case $Z(l) > Z_{max}$ (equivalently, $l < l_{min}$) when for every segment $b$ of adsorbed length both a negative counterion of the sphere and a positive counterion of the chain are released, while the "+" sign refers to the case $Z(l) < -Z_{max}$ ($l > l_{min}$) when for every adsorbed segment a positive counterion is transferred from the chain to the sphere. The various cases are illustrated in Fig. 1.

We can use Eqs. (8) to (11) to discuss the onset of complexation as a function of chain stiffness (at zero tension). For large $l_P$ there is no complexation and we have the case $Z(l) > Z_{max}$ with $B^- > 0$, implying − see Eq. (10) − that the free energy is minimized for $l = 0$. As we reduce $l_P$, $B^-$ changes sign marking the onset of complexation. This complexation occurs in a discontinous fashion: At $B^- = 0$ the wrapped length "jumps" from $l = 0$ to $l = l_{min}$. Then for $B^- < 0$ one has $l > l_{min}$. More specifically, the position of the free energy minimum $l^*$ is given by:

$$l^* = l_{iso} - AR/\mathbf{x} \qquad (12)$$

according to Eq. (8). As long as $A > 0$ the complex in undercharged. Further reduction of $l_P$ leads to (decreasing $A$ and hence) increasing $l^*$ until the complex reaches the isoelectric point at $A = 0$. For smaller $l_P$, $A < 0$ and, according to Eq. (12), $l^* > l_{iso}$, i.e., the complex is overcharged. Consequently, for a fully flexible chain with $l_P = 0$, the complex is always overcharged if $Z \gg Z_{max}$.

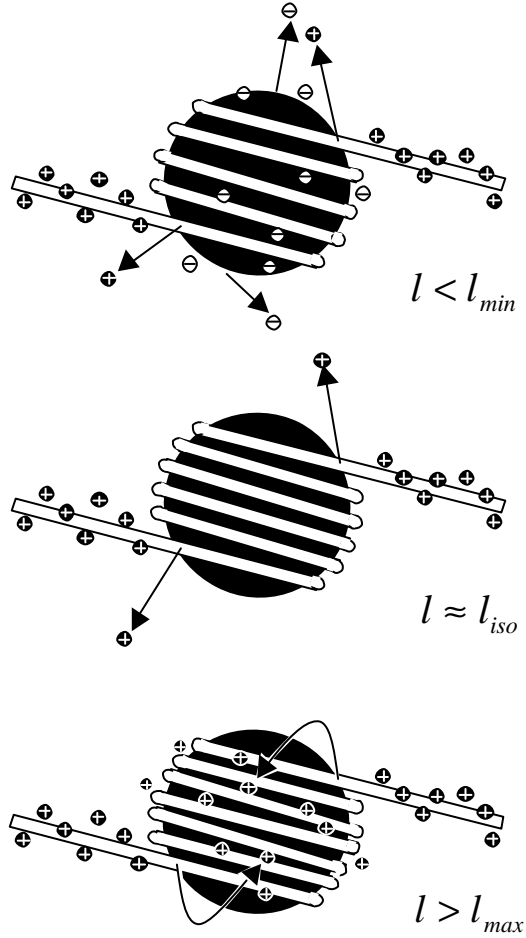

FIG. 1. *Schematic view of the single-sphere/chain complex. The counterions of the positive sphere and negative chain are shown explicitly. Depicted are three scenarios: For $l < l_{min}$ all the counterions of the wrapped part of the chain are released but there is still a fraction of the counterions of the sphere present. By adding a further short piece of chain to the complex, counterions of the sphere and chain are released (arrows). At $l \approx l_{iso}$ there are no negative counterions left on the complex. Addition of chain is driven exclusively by release of counterions of the chain. For wrapping lengths beyond $l_{max}$ there is no further release of counterions; rather the positive counterions of the chain are just transferred to the complex.*

The various regimes can also be traversed as a function of the external tension $f$. This requires replacing $F(l)$ by $F(l) + fl$ and $A$ by $A + fb/k_B T$. For $f = 0$ and sufficiently small $l_P$ one has $l = l^*$ with $l^*$ given by Eq. (12). The corresponding end-to-end distance of the chain is given by $S \cong L - l^* + 2R$ (the exact value of $S$ depends on the location of the points where the chain enters and exits the complex). With increasing $f$ wrapped chain becomes more and more unraveled which leads to an increasing end-to-end distance:

$$S \cong L - l^* + 2R + \frac{bR}{k_B T \chi} f \qquad (13)$$

When the critical force

$$f_{max} = -\frac{k_B T}{b} B^- \qquad (14)$$

is reached the complex becomes unstable and unravels completely, i.e., $l = 0$ and $S = L$. If instead the end-to-end distance $S$ is imposed the force increases linearly with $S$ up to the point $S = L - l_{min}$ where a plateau with $f = f_{max}$ is reached, cf. Fig. 2.

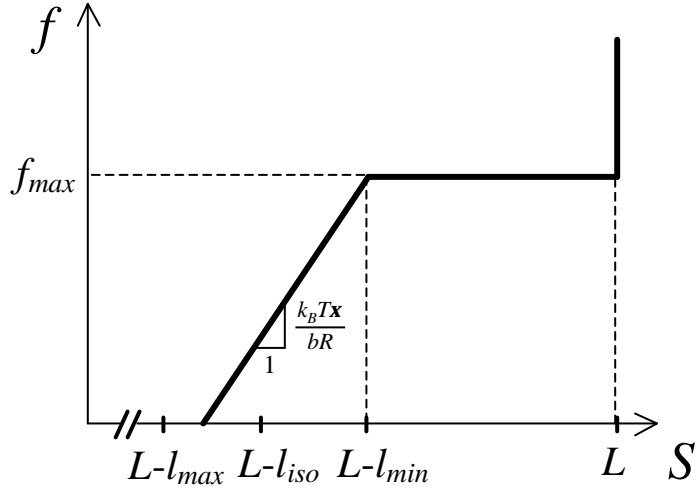

*FIG. 2. Force-extension curve of the single-sphere/chain complex. The linear increase of f with S is due to the charging contribution of the sphere. Further unwrapping leads to the transfer of counterions from the solution to the chain and to the sphere – resulting in a plateau in the stress-strain profile. At $S = L$ the inextensibility of the chain leads to a sharp increase in the force.*

If we identify $l = (\Omega + \tilde{\Omega}) k_B T / b$ with the adhesion energy per unit length arising from counterion release, then Eq. (14) reproduces the result by Marky and Manning.[14] The force extension curve is here, however, more complex than assumed in that study. Also it is important to note that the unwrapping does not necessarily imply a complete dissociation of the complex. Rather, the chain can touch the sphere at one point (see Netz and Joanny[26]) or – if it is sufficiently long compared to its persistence length – it can form many-leaf "rosette" structures.[27] If, for instance, the loops have a diameter of the order of the persistence length then the bending energy per loop is of the order $k_B T$. On the other hand, the gain of adsorption energy per contact follows from counterion release and leads usually to a gain of a few $k_B T$.

## 3 Complexation of a chain with multiple spheres

*A) Chemical equilibrium*

We now turn to the central subject of the paper: the complexation of a chain in a solution of spherical macro-ions. We represent the solution as a reservoir with a concentration $c_m$ of uncomplexed macro-ions. The sphere chemical potential in solution is the sum of the usual ideal solution term and the electrostatic free energy of a spherical macroion with $Z \gg Z_{max}$:

$$\frac{\mu_{sphere}}{k_B T} = \ln(c_m R^3) + Z\tilde{\Omega} \qquad (15)$$

The concentration of the macroions is assumed to be so large that the ideal solution term can be neglected, i.e. $\mu_{sphere}/k_B T \cong Z\tilde{\Omega}$. We will determine the number of spheres that have complexed with the chain by requiring this chemical potential to equal that of the complexed spheres. We assume a beads-on-a-string configuration, with a mean spacing $D$ between spheres. The Euclidian distance between the beginning and the end of the chain will be denoted by $S$. Then $N = S/D$ is the number of complexed spheres. The total chain length $L$ is kept fixed. If $l$ denotes the wrapping length per sphere, then $S$ and $L$ are related by:

$$L \cong Nl + S - 2NR \qquad (16)$$

The Gibbs free energy is now

$$G(N,l) = NF(l) + F_{int}(N,l) - fS(N,l) - \mu_{sphere} N \qquad (17)$$

with $F_{int}$ the interaction between the complexed spheres. For a sphere-sphere spacing $D(N,l) = S(N,l)/N$ small compared to $\kappa^{-1}$ but larger than $2R$, the repulsion is electrostatic and given by (approximately, for $S \gg \kappa^{-1}$)

$$F_{int}(N,l) \cong \Lambda k_B T \frac{N l_B Z^2(l)}{D(N,l)} \qquad (18)$$

for $|Z(l)| < Z_{max}$. The quantity $\Lambda$ is a logarithmic factor of the order $2\ln(\kappa^{-1} N/L)$. Finally, for $D < 2R$ adjacent spheres interact via a strong excluded volume interaction.

We now must minimize $G(N,l)$ with respect to both $N$ and $l$. We always will assume that the parameter $B^- > 0$ so that for the single sphere case the wrapping length is $l^*$ given by Eq. (12). Let us first assume that $N$ is so low that $D \gg \kappa^{-1}$. In that case, $l \cong l^*$. We can lower $G(N,l)$ by increasing $N$. Even when $2R < D < \kappa^{-1}$ it is energetically favorable to keep adding spheres to the chain because the chemical potential term $-\mu_{sphere} N$ is larger than the first three terms of $G(N,l)$. Complexation continues until $D \approx 2R$ and the hard-core repulsion term terminates complexation. Since

$D = S(N,l)/N$, it follows from Eq. (16) that the number of spheres $N$ depends on the wrapping length $L$ as:

$$N \cong L/l \qquad (19)$$

This argument holds for any $l_{min} < l < l_{max}$. Using Eq. (19), we obtain a Gibbs free energy $G(N)$ that depends only on $N$, the total number of complexed spheres:

$$\frac{G(N)}{k_B T} \cong N\left\{(\Lambda+1)\frac{l_B}{2Rb^2}(l_{iso} - L/N)^2 - \frac{2Rf}{k_B T} - \frac{m_{sphere}}{k_B T}\right\} + const \qquad (20)$$

The constant term is not dependent on $N$.

Clearly, the first term of Eq. (20) favors the isoelectric configuration $L/N = l_{iso}$. However, because of the second and third terms, we can lower the free energy further by increasing $N$ beyond $L/l_{iso}$. This is not a small effect since $m_{sphere}/k_B T$ is of the order $Z \gg Z_{max}$ while the first term of Eq. (20), the capacitive energy, is of the order $(l_B/R)Z_{max}^2 \approx Z_{max}$ (since $Z_{max} \approx R/l_B$). The spheres in the many-sphere complex are thus undercharged. Physically, we can illustrate this effect by first setting $L/N = l_{iso}$. In that case the complex is isoelectric. Now add one more sphere. By equally redistributing the chain between the $N+1$ spheres, one has a individual wrapping length $l = L/(N+1)$ close to the isoelectric one. Therefore the previously condensed counterions of the added sphere are released and increase their entropy. By adding more and more spheres − while reducing $l = L/N$ − we can liberate more and more counterions.

Minimizing Eq. (20) we find that $G(N)$ has a minimum in the regime $l_{min} < l < l_{max}$ given by:

$$L/N \cong l_{iso}\left(1 - \frac{\tilde{\Omega} + 2Rf/Zk_B T}{\Lambda+1}\frac{R/l_B}{Z}\right) \qquad (21)$$

where we use the fact that $Z \gg Z_{max} \propto R/l_B$. This means that $L/N$ is not much less than $l_{iso}$, so we do remain close to the isoelectric point.

We now can compute the force required to *increase* the number $N$ of spheres by one, starting from $f = 0$:

$$f(N \to N+1) \cong (\Lambda+1)\frac{Zk_B T}{2R}\left(\frac{Z^2}{R/l_B}\right)\frac{b}{L} \qquad (22)$$

Note that from the work of the previous section, the application of a force would be expected to reduce the number of sphere-chain complexes whereas under the present condition of full chemical equilibrium force application now, paradoxically, increases the number of complexes. Note that in the thermodynamic limit of large $L$, the critical tension vanishes as $1/L$.

*B) "Fixed" N*

We assumed in the previous section that the spheres in solution are in chemical equilibrium with the chain. In particular, the rate of spheres "evaporating" from the complex must equal the rate of spheres "condensing" onto the complex. If, in a force extension experiment, we ramp the force such that it is not possible to maintain chemical equilibrium, we would be in a fixed *N* ensemble. We ask the question: what is the optimal number of condensed beads for a given externally imposed strain? Naively, one might expect that the number of beads always is chosen such that each complex is close to its isoelectric wrapping length $l_{iso}$. With increasing end-to-end distance the length $L-S$ available for wrapping decreases and thus the necklace would have to release spheres in order to have the remaining complexed spheres close to the isoelectric point. A consequence of this mechanism would be a sawtooth pattern in the measured force-extension curve.

Let us first consider the chain with all *N* spheres complexed. The free energy follows directly from Eq. (17):

$$F(N,l) = NF(l) + F_{int}(N,l) \qquad (23)$$

We assume in the following the case of equally spaced and well-separated beads for which $L/N$ greatly exceeds $l_{iso}$ and *R*. In the force-free case, $f=0$, we find from Eq. (23) the following optimal wrapping length:

$$l^* \cong l_{iso} - A\frac{R}{x}\left(1 - \frac{2\Lambda RN}{L}\right) \qquad (24)$$

Comparing this result with the single-sphere case, Eq. (12), it can be seen that the extend of over- or undercharging is reduced. By this means the electrostatic repulsion between the complexes is lowered.

We compute now the behavior of the *N*-bead-on-a-string configuration under an imposed end-to-end distance *S*. We assume that the wrapping length is the same for each complex, i.e., it is given by $l(N,S) \cong (L-S+2NR)/N \cong (L-S)/N$. There are two cases: (*i*) For small values of the wrapping length with $l(N,S) < l_{min}$ (and thus $Z(l) > Z_{max}$) there are condensed counterions on the spheres. $F_1$ is then given by Eq. (10) and $F(N,S)$ is linear in *l* (and *S*):

$$\frac{F(N,S)}{k_B T} \cong B - \frac{l}{b}N - N\tilde{\Omega}Z_{max} + N\frac{l_B Z_{max}^2}{2R} + \frac{\Lambda l_B}{L}Z_{max}^2 N^2 + \ln(kR)NZ_{max} \qquad (25)$$

with $l = l(N,S)$. In Eq. (25) we explicitely wrote down the *l*-independent terms that account for the electrostatics of the necklace for $l \to 0$ where $Z(l) = Z_{max}$. (*ii*) For larger wrapping lengths with $l_{min} < l < l_{max}$, all the counterions of the spheres are released. In that case $F_1$ is given by Eq. (8) and the total free energy of the necklace is quadratic in *l* (and *S*):

$$\frac{F(N,S)}{k_B T} \cong \frac{l_B N}{2R} Z^2(l) + A \frac{l}{b} N - N\tilde{\Omega} Z + \Lambda \frac{N^2 l_B Z^2(l)}{L - N(l - 2R)} + \ln(kR) N Z(l) \quad (26)$$

Note that at $l = l_{min} = l_{iso} - bZ_{max}$ (i.e., at $S = L - l_{min} N$) the two free energies have the same value as well as the same derivative with respect to $S$. Here $Z_{max}$ is of the order $R/l_B \left(\tilde{\Omega} - \ln(kR)\right)(1 - 2\Lambda RN/L)$.

We ask now the question if it is favorable for the necklace to hold on to all its spheres or if − for a given value of $S$ − the necklace can lower its free energy by releasing some of its spheres. Assume a complex with $N - m$ complexed spheres and $m$ free ones. In this case the individual wrapping length is given by $l(N - m, S) \cong (L - S)/(N - m)$. The free energy in this case is given by

$$\tilde{F}(m, N, S) \cong F(N - m, S) - m\tilde{\Omega} Z_{max} + m l_B Z_{max}^2 / 2R \quad (27)$$

Here $F(N - m, S)$ is given by Eq. (25) for $l(N - m, S) < l_{min}$ and by Eq. (26) for $l(N - m, S) > l_{min}$. The last two terms in Eq. (27) describe the free energy of the $m$ free spheres.

Let us first consider the case of high ionic strength where the electrostatic interaction between complexes can be neglected, i.e., the case $\Lambda = 0$. Then the expression for wrapping length $l^*$, Eq. (24), reduces to the single sphere case, Eq. (12) (for $f = 0$). The end-to-end distance of this necklace is given by $S^* \cong L - N(l^* - 2R)$.

We compare now the free energies of necklaces with different numbers of beads for a given externally imposed end-to-end distance $S > S^*$. For simplicity, let us first "switch off" the sphere-chain interaction, i.e., set formally $\ln(kR) = 0$. In this case $\tilde{F}(m, N, S)$ is independent of $m$ as long as the individual wrapping length fulfills $l(N - m, S) < l_{min}$, i.e., as long as one is in the linear regime, Eq. (25). So if we neglect sphere-chain interactions we find that − for a given imposed end-to-end distance $S$ − all $(N - m)$-bead necklaces have the *same* free energy as long as $m$ is sufficiently small, namely $m < N - (L - S)/l_{min}$. Furthermore, as discussed above $\tilde{F}(m, N, S)$ crosses over smoothly to the quadratic regime at $l(N - m, S) = l_{min}$ which corresponds to the end-to-end distance $S_m \cong L - (N - m)b(Z - Z_{max})$. The corresponding free energies are shown in Fig. 3(a). Evidently, for $S < S_m$ the system with $m - 1$ free beads has a lower free energy than the one with $m$ free beads.

From this analysis follows that there is no $S$-value where the release of spheres would lead to a lowering of the free energy − as long as we neglect the electrostatic interactions between the constituents of the complex. It is now easy to show that one breaks the degeneracy discussed above by accounting for the sphere-chain interactions. The release of beads is not favorable since one has to overcome the bead-chain attraction $\ln(kR) Z_{max}$ per released sphere. This picture will also not change if we include the sphere-sphere interaction term since it is of the order $RN/L$ smaller and is thus only a small correction. Therefore the chain always holds on to all of its spheres. The corresponding free energies as a function of the externally imposed end-to-end distance $S$ are depicted in Fig. 3(b).

Consider now the $N$-bead necklace that has an unperturbed length $S^* \cong L - N(l^*-2R)$ with $l^*$ being the optimal wrapping length per bead, Eq. 24. If we increase $S$ we encounter the following restoring force $f = \partial F(N,S)/\partial S = -N^{-1}\partial F/\partial l$:

$$f \cong \begin{cases} \dfrac{\mathbf{x}k_B T}{RbN}(S - L + N(l^*+2R)) & \text{for } L - N(l^*-2R) \leq S < L - N(l_{min} + 2R) \\ -\dfrac{k_B T}{b} B^- & \text{for } L - N(l_{min} - 2R) \leq S < L \end{cases}$$

For $N = 1$ we recover Eqs. (13) and (14). In the linear regime the slope of the restoring force decreases with increasing $N$. Hence the force that is required to stretch the chain by a given amount $\Delta S$ vanishes in the thermodynamic limit $N \to \infty$ and $L \to \infty$ with $L/N$ fixed.

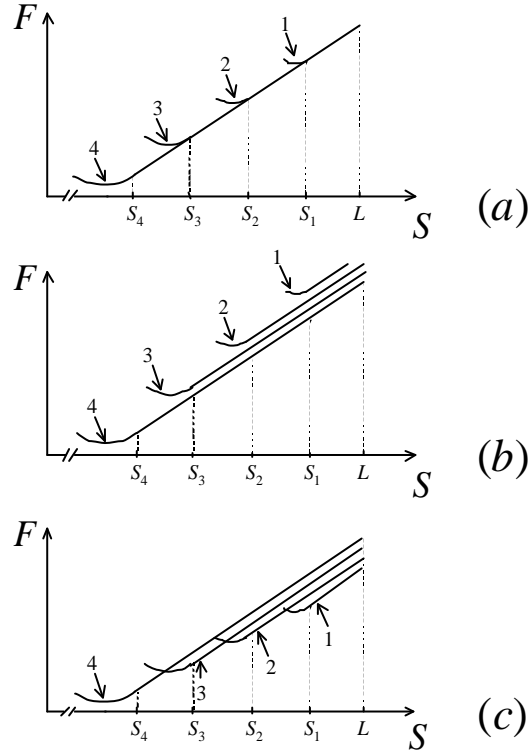

*FIG. 3. Free energies of necklaces with different numbers of beads as a function of the imposed end-to-end distance. Case (a) depicts a special case where necklaces with different numbers of beads are degenerate. As discussed in the text this is found when only the contribution of the counterions is taken into account. (b) The attraction between the spheres and the chain breaks this degeneracy. The chain holds always onto all its spheres. (c) For a large negative contribution to the chemical potential of the spheres (for instance, for a very low density of spheres) the curves intersect and the stretching leads to a sequential release of spheres. In this case one has the occurence of a saw-tooth pattern in the force-extension profile.*

Concluding, we find that the "fixed" *N* case does not lead to release of spheres and sawtooth-like patterns. This is different from a large variety of systems where one has a stepwise unfolding of a chainlike structure under stretching and a sawtooth pattern in the force profile. Experimentally it was observed for the muscle protein titin where the unfolding of domains is responsible for this behavior.[17,18] Other promising candidates are polymers that self-assemble into chains of subunits connected by strings. Polysoaps form strings of micelles.[28] Polyelectrolytes (PEs) in poor solvent,[29] as well as polyampholytes[30,31] (polymers that carry positive and negative charges), assume necklace-type configurations (globules connected by strings). The stepwise unfolding was studied in detail for PEs in a poor solvent[32,33] where a sawtooth pattern in the force profile is predicted. Each step corresponds to the disintegration of a globule and the redistribution of its material between the remaining beads. In this way the necklace can lower its surface energy. Important in these systems is that the subunits have a preferential size: If they can aggregate to one large homogeneous globule, as is the case for a neutral polymer in a poor solvent, one finds just one plateau in the force profile corresponding to the unwinding of the globule.[34]

The chain-necklace discussed in this paper has a preferential wrapping length close to the isoelectric wrapping length $l_{iso}$, that minimizes the charging energy. The mechanism that prevents stepwise unfolding is the free energy loss by releasing a sphere. The contribution of counterion condensation on the sphere is already so large so that one finds the degenerate case depicted in Fig. 3(a). The electrostatic sphere-chain interaction leads to a further shift of the free energies leading to the case depicted in Fig. 3(b). Only if we assume an additional large negative contribution to the chemical potential of the spheres, Eq. (15), it is possible to achieve a situation at which the curves intersect, as depicted in Fig. 3(c). In this case one has a sequential release of the spheres and a sawtooth-pattern in the force-extension profile.

Therefore it cannot be expected that the charging contribution would induce a stepwise unfolding of a beads-on-a-string complex like the chromatin fiber. The sequential release of histone octamers would be too costly due to counterion condensation effects. It is also important to note that for physiological salt concentrations (100*mM*) the charging contributions are effectively negligible due to screening effects ($k^{-1} \approx 10$ Å). However, other factors might be important. For instance, inhomogeneities of the bending properties of DNA due to its base pair sequence lead to different binding energies of the histone octamers. This might lead to their sequential release. Recent microrheological stretching experiment[35] on single chromatin fibers show indeed a non-reversible increase of the fiber length when the fiber is stretched up to a point where the restoring force is of the order of 20*pN*. Subsequent stretch-release cycles find then further non-reversible increase in the stretching length at higher and higher forces (up to $\approx 50 pN$). It is assumed that the irreversible increase in fiber length is due to the loss of histone octamers; the increase in the critical force from cycle to cycle might indicate variations in the binding energy per histone.

# 4 Conclusion

The analytical study presented here shows that a system of polyelectrolytes and oppositely charged spheres can form under- or overcharged complexes depending on the chain flexibility, the concentration of spheres etc. For the case of a single sphere-chain complex we find overcharging for a sufficiently flexible chain – in accordance with a recent study.[10] With increasing chain stiffness (or increasing externally imposed tension) the wrapping length decreases and so the degree of overcharging. By this means it is even possible to have an undercharged complex – up to a critical stiffness (or tension) where an abrupt complete unwrapping of the chain occurs. The structures beyond this unwrapping transition are open multi-leafed "rosettes" that were considered in a prior study.[27]

On the other hand, if the chain is placed in a solution with a finite concentration of spheres the resulting complex is a chain completely "decorated" with spheres, each of which is undercharged. This is the case even for highly flexible chains. This profound difference between single-sphere complexes and multi-sphere complexes is also reflected in the response of such a structure to an externally imposed tension. The single-sphere complex will unwrap gradually and then unwrap abruptly when a tension of the order $k_BT$ per monomer length $b$ is reached. On the other hand, for multisphere complexes applying a tension leads to the surprising effect of the complexation of more and more spheres.

We also considered chains complexed with a "fixed" number $N$ of spheres that show a very soft stretching modulus proportional to $N^{-1}$. Interestingly, the chain holds on to all its spheres up to the point when the critical tension for unwrapping is reached where all spheres unwrap simultaneously.

There is a large amount of work on the problem of complexation of a chain with a single sphere, including theoretical studies[10,14,26,27,36,37,38,39,40,41] and computer simulations.[41,42,43,44] It is important to note that most of these studies find the phenomenon of overcharging, even though all but one[10] do not consider counterion condensation, i.e., these studies are restricted to weakly charged systems. Nguyen and Shklovskii showed in a recent study[41] that the overcharging of these complexes is driven by a correlation effect. For perfectly flexible chains as considered in their study the chain winds around the sphere so that neighboring turns lie parallel at a distance $\Delta$ of the order $R^2/l$. The interaction of the chain with itself beyond $\Delta$ is effectively screened leading to a decrease of the self-energy of the polyelectrolyte upon adsorption. This argument has to be somewhat modified for semiflexible chains where the resulting path of the wrapped chain is more complicated and – at least for short wrapping lengths – resembles a tennisball seam pattern (cf., for instance, Ref. 40).

The above mentioned correlation effect can be neglected, i.e., the charges of the adsorbed chain can be simply "smeared out" on the sphere if the wrapping of the chain around the sphere is sufficiently tight, $\Delta \approx r$; accordingly, for simplicity, the present study does not consider correlation effects. In the opposite case, when $\Delta \gg r$, correlation effects have to be carefully taken into account. At the same time the counterion release mechanism becomes less important. This was shown by Sens and Joanny[13] for the case of the adsorption of a highly charged rod on an oppositely charged planar surface; the fraction of counterions released from the rod upon its adsorption decreases with decreasing surface charge density of the plane.

There are two recent studies that are also devoted to multisphere adsorption on a polyelectrolyte. Jonsson and Linse[45] performed Monte Carlo simulations of a flexible chain complexing with oppositely charged spheres, accounting explicitly for the counterions of the chain and the spheres. Their findings show the same qualitative features concerning over- and undercharging that are found in the present study. A single sphere is usually overcharged by the complexed part of the chain, whereas in the case of an abundance of spheres the number of complexed spheres exceeds the number that is required to form an isoelectric complex, i.e., each bead is undercharged. An analytical approach to the multisphere complex was given by Nguyen and Shklovskii[16] extending their single-sphere theory.[41] Again, for their system also (flexible polyelectrolytes and spheres with no counterion condensation) they find similar features. For a small number of complexed spheres each bead is overcharged by the chain (due to the correlation effect) whereas in the opposite limit the chain will be undercharged. The authors also computed the response of their system to an externally imposed strain: for sufficiently strong screening more and more beads associate with the chain with increasing tension. Interestingly, for the weak screening case it is found that complexed beads leave the chain one by one under increasing tension – reminiscent of the behavior of a polyelectrolyte necklace in a poor solvent.[32,33]

*Acknowledgement*: We would like to thank M. Jonsson, K.-K. Kunze, T. T. Nguyen and B. I. Shklovskii for useful discussions. This work was supported by the National Science Foundation under Grant DMR-9708646.